\documentclass[reqno]{article}

\usepackage{amsmath,amssymb}
\usepackage{latexsym}
\usepackage{cite}
\usepackage{graphicx}
\usepackage{epsfig}
\usepackage{fancyheadings}

\newcommand{\beqs}{\begin{equation*}}
\newcommand{\beq}{\begin{equation}}

\newcommand{\eeqs}{\end{equation*}}
\newcommand{\eeq}{\end{equation}}

\newcommand{\beqas}{\begin{eqnarray*}}
\newcommand{\beqa}{\begin{eqnarray}}

\newcommand{\eeqas}{\end{eqnarray*}}
\newcommand{\eeqa}{\end{eqnarray}}

\newcommand{\eps}{\varepsilon}
\newcommand{\al}{\alpha}

\newcommand{\om}{\omega}

\newcommand{\Om}{\Omega}

\newcommand{\blist}{\begin{itemize}}

\newcommand{\elist}{\end{itemize}}

\providecommand{\href}[2]{#2}

\DeclareFontFamily{OT1}{rsfs}{}
\DeclareFontShape{OT1}{rsfs}{m}{n}{ <-7> rsfs5 <7-10> rsfs7 <10->rsfs10}{} 
\DeclareMathAlphabet{\mycal}{OT1}{rsfs}{m}{n}

\newcommand{\gthree}{{\gamma_\ast}}

\newcommand{\cc}{^*}

\newcommand{\jq}{{j_q}}
\newcommand{\jp}{{j_p}}

\newcommand{\matp}{\mathfrak{p}}
\newcommand{\matq}{\mathfrak{q}}

\newcommand{\ve}{\varepsilon}

\newcommand{\diff}[1][]{\mbox{d}#1}

\newcommand{\gf}[1]{\itindex{#1}{gf}}
\newcommand{\half}[1]{\ensuremath{\frac{#1}{2}}}
\newcommand{\intd}[1]{\int \!\! #1 \;}
\newcommand{\inv}[1]{\ensuremath{\frac{1}{#1}}}

\newcommand{\Stext}[1]{\itindex{\mathcal{S}}{#1}}

\newcommand{\derfrac}[2][]{\frac{\partial #1}{\partial #2}}

\newcommand{\itindex}[2]{\ensuremath{#1_{\mbox{\scriptsize{\itshape #2}}}}}

\DeclareMathOperator{\extdm}{d}
\newcommand{\extd}{\extdm \!}

\title{Quantum dilaton supergravity in 2D \\ with non-minimally
  coupled matter}

\bigskip
\author{Luzi Bergamin \\
\small\it Institute for Theoretical Physics,\\[-1.mm]
\small\it Vienna University of Technology, 1040 Vienna, Austria,  \\[-1.mm]
\small\it email: bergamin@tph.tuwien.ac.at}
\date{ }

\begin{document}

\maketitle
\thispagestyle{fancy}

\rhead{TUW-04-22\\hep-th/0408229}

\bigskip

 \begin{abstract}
 General $N=(1,1)$ dilaton supergravity in two dimensions allows a 
 background independent exact quantization of the geometric part, 
 if these theories are formulated as  specific graded Poisson-sigma 
 models. In this work the extension of earlier results to models with
 non-minimally coupled matter is presented. In particular, the
 modifications of the constraint algebra due to non-minimal couplings
 are calculated and it is shown that quartic ghost-terms do not
arise. Consequently the path-integral quantization as known from
bosonic theories and supergravity with minimally coupled matter can be
taken over.


 \end{abstract}

\section{Introduction}
 The traditional formulation of $N=(1,1)$ dilaton supergravity 
 models in two dimensions is based upon superfields
 \cite{Howe:1979ia,Park:1993sd}. Beside other problems, this approach
 has serious limitations concerning the quantum theory: Superspace
 techniques are known for the second order formulation with vanishing
 bosonic torsion, only. However, as is known from the bosonic case
 (refs.\ \cite{Grumiller:2002nm,Grumiller:2003sk} and refs.\ therein), a
 background independent quantization must start from a first order
 formulation, which in general works with non-vanishing bosonic
 torsion. This formulation can be interpreted as a Poisson-sigma model (PSM) \cite{Schaller:1994es,Schaller:1994uj} of 
 gravity where beside the zweibein also the spin
 connection appears as an \emph{independent} variable 
 together with new scalar fields (``target-space coordinates'') 
 on the 2D world sheet. Upon elimination of certain fields this
 formulation is locally and globally equivalent to the well-known
 second order formulation \cite{Katanaev:1996bh}.

During the last years it has been shown that \emph{graded}
 Poisson-sigma models (gPSMs) provide a natural extension to describe
 dilaton supergravity, if the 
 target space is extended by a ``dilatino'' and the gauge fields 
 comprise also the gravitino
 \cite{Izquierdo:1998hg,Ertl:2000si}. Among the generic class of gPSM
 gravity models a certain subclass of
 ``genuine'' dilaton supergravity  has been identified in
 \cite{Bergamin:2002ju,Bergamin:2003am}, which was shown to be
 equivalent to the model of ref.\ \cite{Park:1993sd} when certain 
 components of the superfield in the latter are properly 
 expressed in terms of the fields appearing in the gPSM
 formulation. This equivalence also allowed to derive in a
 straightforward way
  minimal and non-minimal interactions with matter at the gPSM level
 \cite{Bergamin:2003mh}.

All these considerations at the \emph{classical} level provide the
basis for a background independent quantization, analogous to the
bosonic case. As shown in ref.\ \cite{Bergamin:2004us}, this program
indeed can be taken over to the case of supergravity, although
considerable computational complications arise. The result of
\cite{Bergamin:2004us} is restricted to minimally coupled matter and
it is the purpose of the present work to extend it to
non-minimal coupling.

 This paper is organized as follows: section \ref{sec:two} briefly reviews the
 gPSM formulation of supergravity and the coupling to
 matter fields. In section \ref{sec:constraints} a Hamiltonian analysis is
 performed, whereby emphasis is placed on the difference to the case
 of minimal coupling. This is the necessary prerequisite for the construction of the BRST
 charge and the path integral quantization (sect.\ \ref{sec:3}). Finally, appendix
 \ref{sec:notation} summarizes our notations and conventions.

\section{Graded Poisson-Sigma model and supergravity}
\label{sec:two}
A general gPSM consists of scalar fields
$X^I(x)$, which are themselves (``target space'') coordinates of a graded Poisson manifold with
Poisson tensor $\{X^I,X^J\} = P^{IJ}(X) = (-1)^{IJ+1} P^{JI}(X)$. The index
$I$, in the generic case, may include commuting as well as anti-commuting
fields. In addition one introduces the gauge
potential $A = \extd X^I A_I = \extd X^I A_{mI}(x) \extd x^m$, a one form with respect to the Poisson
structure as well as with respect to the 2D worldsheet. The gPSM
action reads
\begin{equation}
  \label{eq:gPSMaction}
  \begin{split}
    \Stext{gPSM} &= \int_{\mathcal{M}} \extd X^I \wedge A_I + \half{1} P^{IJ}
    A_J \wedge A_I\ . 
  \end{split}  
\end{equation}
As the Poisson tensor $P^{IJ}$ must have vanishing Nijenhuis tensor the action \eqref{eq:gPSMaction} is invariant under the
symmetry transformations $\delta X^{I} = P^{IJ} \ve _{J}$ , $\delta A_{I} = -\mbox{d} \ve
  _{I}-\left( \partial _{I}P^{JK}\right) \ve _{K}\, A_{J}$,
where the term $\extd \epsilon_I$ in the latter relation provides
the justification for calling $A_I$ ``gauge fields''.
If the Poisson
tensor has a non-vanishing kernel---the actual situation in any application to 2D
(super-)gravity due to the odd dimension of the bosonic part of the tensor---there exist (one or more) Casimir functions $C(X)$ obeying
$\{ X^I, C \} = 0$,
which, when determined by the field equations,
are constants of
motion.

In the most immediate application to 2D supergravity the gauge
potentials comprise the spin connection $\om^a{}_b=\eps^a{}_b\om$, the dual basis $e_a$ containing the zweibein and the gravitino $\psi_\al$:
\begin{align}
  A_I &= (A_\phi, A_a, A_\alpha) = (\omega, e_a, \psi_\alpha) & X^I &= (X^\phi,
  X^a, X^\alpha) = (\phi, X^a, \chi^\alpha)
\end{align}
The fermionic components \( \psi _{\alpha } \) (``gravitino'') and \( \chi
^{\alpha } \) (``dilatino'') for $N=(1,1)$ supergravity are Majorana
spinors. The scalar field $\phi$ will be referred to as
``dilaton''. The remaining bosonic target space coordinates $X^a$
correspond to directional derivatives of the dilaton in the second
order formulation (cf.\ \cite{Ertl:2000si,Bergamin:2003am}). Not any
gPSM with this field content can be interpreted as supergravity.
Local Lorentz invariance determines the $\phi$-components of the Poisson
tensor
\begin{align}
\label{eq:lorentzcov}
  P^{a \phi} &= X^b {\epsilon_b}^a\ , & P^{\alpha \phi} &= -\half{1}
  \chi^\beta {\gthree_\beta}^\alpha\ ,
\end{align}
and the supersymmetry transformation is encoded in $P^{\alpha
  \beta}$. Symmetry restrictions of the latter
  \cite{Bergamin:2002ju,Bergamin:2003am} yields 
 the gPSM \emph{supergravity}
class of theories (called ``minimal field supergravity'', MFS, in our present paper) with the Poisson
tensor (\( \chi ^{2}=\chi ^{\alpha }\chi _{\alpha }\), $Y = X^a X_a
  / 2$ and $f'$ denotes $d f /d \phi$)
\begin{gather}
\label{eq:mostgensup}
  P^{ab} = \biggl( V + Y Z - \half{1} \chi^2 \Bigl( \frac{VZ + V'}{2u} +
  \frac{2 V^2}{u^3} \Bigr) \biggr) \epsilon^{ab}\ , \\
\label{eq:mostgensuplast}
\begin{alignat}{2}
  P^{\alpha b} &= \frac{Z}{4} X^a
    {(\chi \gamma_a \gamma^b \gthree)}^\alpha + \frac{i V}{u}
  (\chi \gamma^b)^\alpha\ , \qquad
  P^{\alpha \beta} &= -2 i X^c \gamma_c^{\alpha \beta} + \bigl( u +
  \frac{Z}{8} \chi^2 \bigr) \gthree^{\alpha \beta}\ ,
\end{alignat}
\end{gather}
where the three functions $V$, $Z$ and the
``prepotential'' $u$ depend on the
dilaton field $\phi$ only and must be related by
$V\left( \phi \right) =- \inv{8} \bigl(( u^{2})' + u^{2} Z\left( \phi
\right) \bigr)$.
Inserting the Poisson tensor \eqref{eq:lorentzcov},
\eqref{eq:mostgensup} and \eqref{eq:mostgensuplast}
into equation \eqref{eq:gPSMaction} the ensuing action becomes (for
simplicity the wedge symbols are omitted, $D$ denotes the covariant
derivatives  $(D e)_a = \extd e_a + \omega {\epsilon_a}^b e_b$, $(D \psi)_\alpha = \extd
  \psi_\alpha - \half{1} {{\omega \gthree}_\alpha}^\beta \psi_\beta$)
\begin{multline}
  \label{eq:mostgenaction}
  \Stext{MFS} = \int_{\mathcal{M}} \bigl( \phi \diff \omega + X^a D e_a + \chi^\alpha D
  \psi_\alpha + \epsilon \biggl( V + Y Z - \half{1} \chi^2 \Bigl( \frac{VZ + V'}{2u} +
  \frac{2 V^2}{u^3} \Bigr) \biggr) \\
   + \frac{Z}{4} X^a
    (\chi \gamma_a \gamma^b e_b \gthree \psi) + \frac{i V}{u}
  (\chi \gamma^a e_a \psi) 
   + i X^a (\psi \gamma_a \psi) - \half{1} \bigl( u +
  \frac{Z}{8} \chi^2 \bigr) (\psi \gthree \psi) \bigr)\ .
\end{multline}
The Poisson tensor has at least one (bosonic) Casimir function $C = e^Q \bigl(Y- u^2/8 + \chi^- \chi^+ ( u' + \half{1} u Z)/8\bigr)$.
In certain situations additional (bosonic and fermionic) Casimir
functions emerge
\cite{Ertl:2000si,Bergamin:2003am}.

It has been proven in \cite{Bergamin:2003am} that this class of supergravity models is equivalent to the superfield supergravity of Park and Strominger
\cite{Park:1993sd} upon elimination of auxilliary fields and a suitable redefinition of
the gravitino.
This equivalence can be used to derive from the superspace construction the
matter coupling for MFS models, for details of the calculations we
refer to \cite{Bergamin:2003mh}. A supersymmetric matter multiplet consists of
a real scalar field $f$ and a Majorana spinor $\lambda_\alpha$. In case of
non-minimal coupling a coupling function $K(\phi)$ is introduced as
well. After elimination of all auxiliary fields the matter action 
\begin{multline}
\label{eq:cmMFS}
      \mathcal{S}_{(m)} = \int_{\mathcal{M}}  \biggl[ {K}  \Bigl( \half{1} \extd {f} \wedge
    \ast \extd {f} + \half{i}
    {\lambda} \gamma_a {e}^a \wedge \ast \extd {\lambda} + i \ast({e}_a \wedge \ast \extd {f}) {e}_b \wedge \ast {\psi} \gamma^a
    \gamma^b {\lambda}
    + \inv{4} \ast ({e}_b \wedge \ast {\psi}) \gamma^a \gamma^b
    {e}_a \wedge \ast {\psi} {\lambda}^2 \Bigr) \\ +
    \frac{{u}}{8} {K}' {\lambda}^2 {\epsilon} - \inv{4}
    {K}'  ({\chi} \gthree \gamma^a {\lambda}) {e}_a \wedge \ast
    \extd {f} - \inv{32}\Bigl({K}''  - \half{1}
    \frac{\bigl[{K}' \bigr]^2}{{K}}\Bigr) {\chi}^2 {\lambda}^2 {\epsilon} \biggr]
\end{multline}
is found to be invariant under local non-linear supersymmetry
transformations (cf.\ eqs.\ (2.12)-(2.17) and (6.23)-(6.27) of \cite{Bergamin:2003mh}).

\section{Hamiltonian analysis}
\label{sec:constraints}
The primary goal of the present paper is to develop the systematics of the
quantization of the action
\eqref{eq:mostgenaction} together with matter couplings \eqref{eq:cmMFS}.
The quantization is performed via a Hamiltonian analysis introducing Poisson
brackets. The special case of minimal coupling ($K(\phi) \equiv
1$) was addressed in \cite{Bergamin:2004us}. Here we concentrate onto the
difficulties arising from a non-trivial coupling function $K$, for
further details on the purely geometric models and minimal coupling refs.\
\cite{Bergamin:2004us,Bergamin:2002ju} as well as the literature on non-supersymmetric models 
(ref.\ \cite{Grumiller:2002nm} and refs.\ therein)
should be consulted.

In what follows quantities evaluated from the gPSM part of the
action are indicated by $(g)$, quantities from the matter extension by $(m)$.  This separation is possible as the matter action
\eqref{eq:cmMFS} does not contain derivatives acting on the MFS
fields. Further all formulae are written in light-cone coordinates
from know on.

In the geometrical sector we define the canonical variables and the first class primary
constraints ($\approx$ means zero on the surface of constraints) from the Lagrangian $L_{(g)}$ in \eqref{eq:gPSMaction}
($\dot{q}^I = \partial_0 X^I$) by
\begin{align}
\label{eq:gaugepot}
  X^I &= q^I\ , & \bar{q}^I &= (-1)^{I+1} \derfrac[L]{\dot{\bar{p}}_I} \approx 0\ ,
   &\derfrac[L]{\dot{q}^I} &= p_I = A_{1 I}\ , & \bar{p}_I &= A_{0 I}\ .
\end{align}
From the Hamiltonian density ($\partial_1 = \partial$)
\begin{equation}
\label{eq:hamdensity}
  H_{(g)} =  \dot{q}^I p_I - L_{(g)} = \partial q^I \bar{p}_I - P^{IJ} \bar{p}_J p_I
\end{equation}
the graded canonical equations $\partial H_{(g)}/\partial {p_I} =
  (-1)^I \dot{q}^I$ and $\partial H_{(g)}/ \partial {q^I} = -
  \dot{p}_I$
are consistent with the graded Poisson
  bracket for field monomials $\{q^I, p'_J\} = (-1)^I \delta^I_J \delta(x-x')$.
The primes\footnote{Derivatives with respect to the dilaton are
  indicated by a prime as well. We leave this inconvenience of
  notation as dots are used to indicate derivatives with respect to $x^0$.} indicate the dependence on primed world-sheet
coordinates $x$, resp.\ $x'$, $x''$. The Hamiltonian density
\eqref{eq:hamdensity} $H_{(g)} = G_{(g)}^I \bar{p}_I$
is expressed in terms of secondary constraints only:
\begin{equation}
\label{eq:gconstraints}
  \{\bar{q}^I, \intd{\diff{x^1}} H_{(g)} \} = G_{(g)}^I = \partial q^I + P^{IJ} p_J 
\end{equation}

The extension to include conformal matter is straightforward. From the action
\eqref{eq:cmMFS} together with the matter fields $\matq = f$,
$\matq^\alpha = \lambda^\alpha$ 
the canonical momenta\footnote{In what follows, canonical variables
  are used throughout. $\sqrt{-g}=e
= p_{--} \bar{p}_{++} - p_{++} \bar{p}_{--}$ in these variables.}
\begin{gather}
  \begin{split}
\label{eq:bosonP}
    \derfrac[L_{(m)}]{\dot{\matq}} = \matp&= \frac{K}{e} \Bigl( ( p_{++} \bar{p}_{--} + p_{--} \bar{p}_{++} ) \partial \matq -
    2 p_{++} p_{--} \dot{\matq} + 2 i \bigl(p_{++} p_{--} (\bar{p}_- \matq^- + \bar{p}_+
    \matq^+) \\ &\quad - \bar{p}_{++} p_{--} p_+ \matq^+ - p_{++} \bar{p}_{--}
    p_- \matq^- \bigr) \Bigr)  + \frac{i}{2 \sqrt{2}} K' (p_{++} q^+ \matq^+ + p_{--} q^- \matq^-)\ ,
  \end{split}\\
\label{eq:P+}
\begin{alignat}{2}
\derfrac[L_{(m)}]{\dot{\matq}^+} =  \matp_+ &= - \frac{K}{\sqrt{2}}
p_{++} \matq^+\ , \qquad
\derfrac[L_{(m)}]{\dot{\matq}^-} = \matp_- &= \frac{K}{\sqrt{2}} p_{--}  \matq^-
\end{alignat}
\end{gather}
are obtained. Analogous to \eqref{eq:hamdensity} the Hamiltonian density from
the matter Lagrangian in eq.\ \eqref{eq:cmMFS} is defined as $H_{(m)} = \dot{\matq} \matp + \dot{\matq}^+ \matp_+ + \dot{\matq}^- \matp_- - L_{(m)}$,
and the total Hamiltonian density is the sum of this contribution and
\eqref{eq:hamdensity}. For the Poisson bracket of two matter field monomials one finds
 $\{ \matq, \matp' \} = \delta(x - x')$ and  $ \{ \matq^\alpha, \matp'_\beta\} = -
  \delta^\alpha_\beta  \delta(x - x')$.
We do not provide the explicit form of the matter
Hamiltonian, as it can again be written in terms of secondary constraints:
\begin{align}
\label{eq:classicalham}
  H &= G^I \bar{p}_I & G^I &= G^I_{(g)} + G^I_{(m)} & \{\bar{q}^I, \intd{\diff{x^1}} H_{(m)} \} = G_{(m)}^I
\end{align}
The explicit expressions for the matter part of the secondary constraints read:
\begin{align}
\label{eq:G++m}
\begin{split}
  G^{++}_{(m)} &= - \frac{K}{4 p_{++}} (\partial \matq - \inv{K} \matp)^2 +
  i (\partial \matq - \inv{K} \matp) (\frac{K}{p_{++}} p_+
  \matq^+  - \frac{K'}{4 \sqrt{2}} \frac{p_{--}}{p_{++}} q^- \matq^-) +  \frac{i K'}{4 \sqrt{2}}  (\partial \matq + \inv{K}
   \matp) q^+ \matq^+ \\ &\quad + \frac{K}{\sqrt{2}} \matq^+ \partial \matq^+ -
   \frac{K'}{2 \sqrt{2}} \frac{p_{--}}{p_{++}} p_+ q^- \matq^- \matq^+ - p_{--} \Bigl( \frac{u K'}{4} - \inv{8} (K'' -
   \frac{K'^2}{K}) q^- q^+ \Bigr) \matq^- \matq^+
\end{split}
  \\
\label{eq:G--m}
\begin{split}
  G^{--}_{(m)} &=  \frac{K}{4 p_{--}} (\partial \matq + \inv{K} \matp)^2 -
  i (\partial \matq + \inv{K} \matp) (\frac{K}{p_{--}} p_-
  \matq^-  + \frac{K'}{4 \sqrt{2}} \frac{p_{++}}{p_{--}} q^+ \matq^+) +  \frac{i K'}{4 \sqrt{2}}  (\partial \matq - \inv{K}
   \matp) q^- \matq^- \\ &\quad- \frac{K}{\sqrt{2}} \matq^- \partial \matq^- +
   \frac{K'}{2 \sqrt{2}} \frac{p_{++}}{p_{--}} p_- q^+ \matq^- \matq^+ + p_{++} \Bigl( \frac{u K'}{4} - \inv{8} (K'' -
   \frac{K'^2}{K}) q^- q^+ \Bigr) \matq^- \matq^+
\end{split}
  \\
\label{eq:G+m}
  G^+_{(m)} &= i K (\partial \matq - \inv{K}\matp) \matq^+ -
  \frac{K'}{2 \sqrt{2}} p_{--} q^- \matq^- \matq^+ \\
\label{eq:G-m}
  G^-_{(m)} &= -i K (\partial \matq + \inv{K}\matp) \matq^- +
  \frac{K'}{2 \sqrt{2}} p_{++} q^+ \matq^- \matq^+
\end{align}

As the kinetic term of the matter fermion $\lambda$ is first order only, this
part of the action leads to constraints as well. From \eqref{eq:P+} the usual primary
second-class constraints are deduced:
\begin{align}
\label{eq:Psiconstr+}
   \Psi_+ &= \matp_+ + \frac{K}{\sqrt{2}} p_{++} \matq^+ \approx 0 &
   \Psi_- &= \matp_- - \frac{K}{\sqrt{2}} p_{--} \matq^- \approx 0
\end{align}
These second class constraints are treated by substituting the Poisson bracket
by the ``Dirac bracket'' \cite{Dirac:1996} $\{ f, g \}^* = \{ f, g \} - \{f, \Psi_\alpha\} C^{\alpha \beta} \{\Psi_\beta,
  g\}$, where $C^{\alpha \beta} C_{\beta \gamma} = \delta^\alpha_\gamma$ and $C_{\alpha \beta}
  = \{\Psi_\alpha, \Psi_\beta\}$.
From \eqref{eq:Psiconstr+} together with the
definition of the canonical bracket the matrix $C_{\alpha \beta}$
follows as $C_{\alpha \beta} = \sqrt{2} K \mbox{diag}( - p_{++},p_{--})$
The Dirac brackets among the fermionic matter variables are:
\begin{align}
\label{eq:dirbr1}
  \{\matq^+, {\matq'}^+ \}^* &= \inv{\sqrt{2} K p_{++} } \delta(x - x') & \{\matq^-, \matq^- \}^* &=
  - \inv{\sqrt{2} K p_{--}} \delta(x - x')  \\
\label{eq:dirbr2}
  \{\matp_+, \matp'_+\}^* &= \frac{K p_{++}}{2 \sqrt{2}} \delta(x - x') & \{\matp_-,
  \matp'_-\}^* &= -\frac{K p_{--}}{2 \sqrt{2}} \delta(x - x') \\
\label{eq:dirbr3}
  \{\matq^+,\matp'_+\}^* &= - \half{1} \delta(x - x') & \{\matq^-,\matp'_-\}^* &= - \half{1}
  \delta(x - x')
\end{align}
Moreover, $q^{++}$, $q^{--}$ and $p_\phi$ have non-trivial Dirac
brackets with the matter fermion:
\begin{align}
\label{eq:dirbr4}
  \{q^{++}, \matp'_+\}^* &= - \frac{K}{2\sqrt{2}} \matq^+ \delta(x - x') & \{q^{--},
  \matp'_-\}^* &= \frac{K}{2\sqrt{2}} \matq^- \delta(x - x') \\
\label{eq:dirbr5}
  \{q^{++}, {\matq'}^+\}^* &= - \frac{\matq^+}{2 p_{++}} \delta(x - x') & \{q^{--},
  {\matq'}^-\}^* &= - \frac{\matq^-}{2p_{--}} \delta(x - x') \\
\label{eq:dirbr6}
  \{p_\phi, \matp'_+ \}^* &= \frac{K'}{2 \sqrt{2}} p_{++} \matq^+ \delta(x-x') &
  \{p_\phi, \matp'_- \}^* &= - \frac{K'}{2 \sqrt{2}} p_{--} \matq^- \delta(x-x')
  \\
\label{eq:dirbr7}
  \{p_\phi, {\matq'}^+\}^* &= \frac{K'}{2 K} \matq^+
  \delta(x-x') & \{p_\phi, {\matq'}^-\}^* &= \frac{K'}{2 K} \matq^-
  \delta(x-x')
\end{align}
All remaining brackets are unchanged.

An
important step is the calculation of the algebra of secondary
constraints
\begin{align}
   \label{eq:constralg} \{ G^I, {G'}^J \}^* &=  G^K {C_K}^{IJ} \delta(x
   - x')\  .
\end{align}
For the matterless case the calculation has been
performed in \cite{Bergamin:2002ju}, the extension with minimally
coupled matter in \cite{Bergamin:2004us}. For all details of this part of the
calculation we refer the reader to these two publications and simply
state the result
\begin{equation}
  \label{eq:new1}
  \itindex{C}{min}{}_K{}^{IJ} = - \partial_K P^{IJ}\ ,
\end{equation}
where the index ``min'' stands for minimal coupling. For the structure
functions of the full, non-minimally coupled theory the notation ${C_K}^{IJ} =  \itindex{C}{min}{}_K{}^{IJ} + \Delta {C_K}^{IJ}$
is useful. Up to total derivatives\footnote{Of course the appearance of
total derivatives suggests the possibility of central charge
extensions of the algebra. Certainly this directions should be
investigated, but this is not the aim of the present work.} new contributions $\Delta {C_K}^{IJ}$ can then be
summarized as follows:
\begin{gather}
\label{eq:new3.1}
\begin{alignat}{2}
  \Delta {C_\phi}^{++|-} &= p_{--} \derfrac{p_-} \Delta
  {C_\phi}^{++|--} &\qquad \Delta {C_\phi}^{--|+} &= - p_{++} \derfrac{p_+} \Delta
  {C_\phi}^{++|--}\end{alignat} \\
  \label{eq:new3.2}
\begin{alignat}{2}
  \Delta {C_+}^{++|--} &= - \inv{2 \sqrt{2}} \Delta {C_\phi}^{--|+} &\qquad
  \Delta {C_-}^{++|--} &= - \inv{2 \sqrt{2}} \Delta {C_\phi}^{++|-}
\end{alignat} \\
  \label{eq:new3.3}
  \Delta {C_\phi}^{+|-} = p_{++} p_{--} \derfrac{p_+} \derfrac{p_-} \Delta
  {C_\phi}^{++|--} \\
  \label{eq:new3.4}
\begin{alignat}{2}
  \Delta {C_+}^{++|-} &= \inv{2 \sqrt{2}} \Delta {C_\phi}^{+|-} &\qquad
  \Delta {C_-}^{--|+} &= - \inv{2 \sqrt{2}} \Delta {C_\phi}^{+|-}
\end{alignat}
\end{gather}
There remains the definition of $\Delta {C_\phi}^{++|--}$. From a
straightforward but tedious calculation the result
\begin{multline}
  \label{eq:new4}
  \Delta {C_\phi}^{++|--}  = \frac{K'}{4 p_{++} p_{--}} \Bigl( (\partial
  \matq + \inv{K} \matp) (\partial
  \matq - \inv{K} \matp) - 2 i \bigl( p_- \matq^- (\partial \matq
  - \inv{K} \matp) +  p_+ \matq^+ (\partial \matq + \inv{K} \matp)
  \bigr) \Bigr) \\
  + \frac{i}{8 \sqrt{2}} \frac{K'^2}{K} \bigl(\frac{q^-
  \matq^-}{p_{++}}(\partial \matq + \inv{K} \matp) -  \frac{q^+
  \matq^+}{p_{--}}(\partial \matq - \inv{K} \matp)  \bigr)
  + \frac{i}{4 \sqrt{2}} (K'' - \frac{K'^2}{K}) \bigl(\frac{q^-
  \matq^-}{p_{++}}(\partial \matq - \inv{K} \matp) -  \frac{q^+
  \matq^+}{p_{--}}(\partial \matq + \inv{K} \matp)  \bigr)\\
  + \frac{K' p_- p_+}{p_{--} p_{++}} \matq^- \matq^+ + \inv{2
  \sqrt{2}} (K'' - \frac{K'^2}{K}) (\frac{p_-
  \matq^+}{p_{--}} +\frac{ p_+ \matq^-}{p_{++}}) \matq^- \matq^+
  + (\partial_\phi - \frac{K'}{K}) \bigl( \frac{u K'}{4} -
  \inv{8} (K'' - \frac{K'^2}{K}) q^- q^+ \bigr) \matq^-\matq^+
\end{multline}
is obtained. One of the important open question concerns the
interpretation of this result, to enable the application of the
methods developed here in other theories, e.g.\ extended
supergravity. We will comment on that point
again below.

\section{Quantization}
\label{sec:3}
As in the case of minimal coupling \cite{Bergamin:2004us} two pairs of
(anti-)ghosts are introduced for the primary constraints $( b_I,
p_b^I)$ and the secondary constraints $(c_I,  p_c^I)$, resp.
The brackets among them are defined conveniently as 
\begin{align}
  [ b_I, p_b^J ] &= - (-1)^{(I+1)(J+1)} [ p_b^J, b_I ] =  \delta_I^J\ , & [ c_I,
  p_c^J ] &= - (-1)^{(I+1)(J+1)} [ p_c^J, c_I ] =  \delta_I^J\ .
\end{align}
To first order in homological perturbation theory the BRST charge $\Om$ follows straightforwardly:
\begin{equation}
\label{eq:omegadef}
  \Omega = \bar{q}^I b_I + G^I c_I - \half{1} (-1)^I p_c^K \partial_K P^{IJ} c_J c_I
\end{equation}
In ref.\ \cite{Bergamin:2004us} it was found that \eqref{eq:omegadef}
is nilpotent for minimally coupled matter. The aim of this section is
to check this result for the non-minimally coupled case. It is
important to realize that the nilpotency of \eqref{eq:omegadef} is far
from obvious in the present case: The symmetry of the theory is
non-linear and moreover the matter extension is obtained from a
superspace version by (non-trivial) eliminations of auxiliary
fields. It is exactly this last point where quartic ghost terms
usually appear in target-space supergravity.

Therefore it is
inevitable to check the nilpotency of \eqref{eq:omegadef}
carefully. We cannot present the lengthy and tedious calculation in
all details, instead we make some comments on the most important steps
and then present the final result.
\begin{enumerate}
\item Obviously all brackets with the first term in
  \eqref{eq:omegadef} vanish trivially, as $\bar{q}^I$ plays the r\^{o}le
  of a Lagrange multiplier in our approach. In fact, this part of the
  BRST charge simply could be omitted (``minimal phase space'').
\item The remaining terms with zero anti-ghosts vanish by means of the
  constraint algebra \eqref{eq:constralg}, which is nothing than the
  definition of the BRST charge to first order in the homological
  perturbation theory. Nevertheless, there appear some subtleties with
  total derivatives, which are illustrated at hand of two examples:
  \begin{enumerate}
  \item The total derivatives mentioned before eq.\ \eqref{eq:new3.1}
  turn into total derivatives including the ghosts and are again
  dropped. E.g.\ the bracket $\{ G^+, G'^- \}\cc$ generates the total
  derivative $\partial (\matq^+ \matq^-)$, which in the present
  calculation turns into a total derivative $\partial(\matq^+ c_+
  \matq^- c_-)$.
  \item A new subtlety appears as brackets of the form $\{ G^{++}
  c_{++}, G'^{++} c'_{++} \}\cc$ are symmetric and thus do not vanish
  trivially. Due to the statistics of the ghosts the only possible
  terms are proportional to $c_{++} \partial c_{++}$. However all
  terms of this kind are found to cancel. 
  \end{enumerate}
\item The different contributions with one anti-ghost yield a lot of
  messy expressions. But at the end of the day one finds
  that---somehow miraculously---all
  contributions cancel. Again it is important to realize that
  contributions of the form $c_I \partial c_I$ for two anti-commuting
  ghosts must be checked as well. Notice that the $\partial q^I$ part
  of the secondary constraints is important for these terms to
  cancel. As an example one finds that the combination $\partial
  q^{++} + K \matq^+ \partial \matq^+ /\sqrt{2}$ in $G^{++}$ (cf.\ \eqref{eq:G++m})
  cancels all surface contributions from the commutation with
  $\matq^+$ according to \eqref{eq:dirbr1} and
  \eqref{eq:dirbr5}. Also, the inverse powers of $p_I$ are important
  to cancel contributions from the scalar matter fields.
\item Much simpler are contributions with two anti-ghosts. Most terms
  vanish identically as the structure functions from minimal coupling
  do not depend on the $p_I$. Again all contributions cancel.
\end{enumerate}
Putting the pieces together we can conclude that the BRST charge
\eqref{eq:omegadef} is nilpotent. Quartic ghost terms do not appear!
In all steps of the calculation no simple symmetry principle can be
found, which is responsible for this result. Rather an interplay
of the structure of the secondary constraints, of the structure
functions and of the Dirac brackets is relevant. Therefore we are not able to
guess, whether this result will apply to a broader class of models
(such as potentials with self-interaction or extended supergravity) or not.

As the Hamiltonian vanishes on the constraint surface it simply
becomes $\gf{H} = \{ \Omega, \Psi \}$
for some gauge fixing fermion $\Psi$. Adopting the result of
\cite{Bergamin:2004us} the  multiplier
gauge $\Psi =  -i p_c^{++}$ is used\footnote{Notice that
  according to our notation $p_{++}$ is purely imaginary. Also, one
  component ($q^+$, $\matq^-$) of a spinor is real, while the other ($q^-$, $\matq^+$) is imaginary.}. This gauge-fixing entails the
Eddington-Finkelstein form of the bosonic line element and the ensuing
Lagrangian becomes
\begin{equation}
  \label{eq:quant12}
\begin{split}
  \gf{L} &= \dot{\bar{q}}^I\bar{p}_I + \dot{q}^I p_I + \dot{\matq} \matp + \dot{\matq}^\alpha
  \matp_\alpha + p_c^I \dot{c}_I + p_b^I \dot{b}_I - i \partial q^{++}
  - i P^{++|J} p_J  - i (-1)^K p_c^I C_I{}^{++|K} c_K\\
    &\quad + \frac{i}{4} \frac{K}{p_{++}} (\partial \matq - \inv{K} \matp)^2 +
  (\partial \matq - \inv{K} \matp) (\frac{K}{p_{++}} p_+
  \matq^+  - \frac{K'}{4 \sqrt{2}} \frac{p_{--}}{p_{++}} q^- \matq^-) +  \frac{K'}{4 \sqrt{2}}  (\partial \matq + \inv{K}
   \matp) q^+ \matq^+ \\ &\quad - \frac{i}{\sqrt{2}} K \matq^+ \partial \matq^+ +
   \frac{i}{2 \sqrt{2}} K' \frac{p_{--}}{p_{++}} p_+ q^- \matq^-
  \matq^+ + i p_{--} \Bigl( \frac{u K'}{4} - \inv{8} (K'' -
   \frac{K'^2}{K}) q^- q^+ \Bigr) \matq^- \matq^+\ .
\end{split}
\end{equation}
This important and surprising result shall now be used
to formulate the quantum theory in terms of a path integral, as done
for minimal coupling in \cite{Bergamin:2004us} already (cf.\ also the literature on
the bosonic case \cite{Grumiller:2002nm}). However, in a non-linear
theory with structure \emph{functions} rather than structure
constants, the result of the Hamiltonian analysis must be used with
care in a path integral formulation; there may appear ordering
problems from the commutation of operators, which yield additional
terms in the path integral where the variables are treated as
classical fields. In bosonic gravity with non-minimal coupling
\cite{Grumiller:2002nm} and in supergravity with minimal coupling
\cite{Bergamin:2004us} such ordering problems are absent. The important observation
is that $[G^K, {C_K}^{IJ}] = 0$
without any ordering terms (cf.\ Appendix B.2 of
\cite{Bergamin:2004us}). 

As in the simplified version with minimal
coupling possible sources of ordering problems are the appearance of
$\matq^+$ or $1/p_{++}$ together with $q^{++}$. However, it can be
checked by explicit calculations that the new contributions from
non-minimal coupling do not spoil this central relation. This behavior is possible as
in the constraints $G^{++}$ and $G^+$ $\matq^-$ and $p_{--}$ appear in
the combination $\matq^- p_{--}$ only, which is free of ordering terms in
$[q^{--},\matq^- p_{--}]$. Similar observations hold for $\Delta {C_K}^{IJ}$.

Therefore we can plug the result \eqref{eq:quant12} into a generating
functional of Green functions, which has to be integrated over all physical fields and all ghosts. Together with
sources $\mathcal{J} = (\jq_I, \jp^I, J, J_\alpha)$ for the geometrical variables and for
the matter fields resp.\ it reads:
\begin{equation}
  \label{eq:quant13}
    \mathcal{W}[\mathcal{J}] = \int
    \mathcal{D}(q^I,p_I,\bar{q}^I,\bar{p}_I,\matq,\matp,\matq^\alpha,\matp_{\alpha},c_I,p_c^I,b_I,p_b^I) \exp \Bigl( i \intd{\diff{^2 x}} \bigl(\gf{L} + q^I \jq_I + \jp^I
    p_I + \matq J + \matq^{\alpha} J_\alpha \bigr)\Bigr)
\end{equation}
The gauge-fixed Hamiltonian, being independent of $\bar{q}^I$, $\bar{p}_I$, $b_I$ and
$p_b^I$, allows a trivial integration of all these fields. As the remaining
ghosts appear at most bi-linearly in the action they can be integrated
over, which leads to the super-determinant $\mbox{sdet}  {M_I}^J = \mbox{sdet} \bigl( {\delta_I}^J \partial_0 +
  i C_I{}^{++|J} \bigr)$.
The integration of $\matp_\alpha$ by means of the constraint
\eqref{eq:Psiconstr+} is trivial as well, while this is possible for the bosonic momentum
$\matp$ of matter after a quadratic completion. This yields the
effective matter Lagrangian
\begin{multline}
  \label{eq:quant15}
  L_{(m)} = \frac{i}{4 K p_{++}} \tilde{\matp}^2 + i
  K p_{++} \dot{\matq}^2 + \dot{\matq} \bigl(K \partial \matq - 2 i K
  p_+ \matq^{+} + \frac{i}{2 \sqrt{2}} K' (p_{--} q^- \matq^- + p_{++}
  q^+ \matq^+)\bigr) + \frac{K'}{2 \sqrt{2}} \partial \matq q^+ \matq^+ 
  \\ - \frac{K}{\sqrt{2}} p_{++} \dot{\matq}^+ \matq^{+} + \frac{K}{\sqrt{2}} p_{--}
  \dot{\matq}^- \matq^{-} - \frac{i}{\sqrt{2}} K \matq^+ \partial
  \matq^+ + i p_{--} \Bigl( \frac{u K'}{4} - \inv{8} (K'' - \inv{2}
   \frac{K'^2}{K}) q^- q^+ \Bigr) \matq^- \matq^+\ .
\end{multline}
After integration over the shifted variable $\tilde{\matp}$ the
complete Lagrangian is linear in the $p_I$, the determinant $\det K
p_{++}$ can be absorbed by a suitable redefinition of the path
integral measure of $\matq$ and $\matq^\alpha$. Therefore, as in the
bosonic case and in supergravity with minimal coupling, all geometric
variables can be integrated out and one is left with the integration
of the matter variables, which must be treated
perturbatively. Although \eqref{eq:quant15} shares many properties
with the respective results from earlier calculations (bosonic
gravity, supergravity with minimal coupling), an important difference
should be mentioned: As a consequence of the elimination of
(superspace) auxiliary fields the prepotential $u$ appears in the
matter Lagrangian \eqref{eq:quant15} and thus plays the r\^{o}le of
the prepotential of geometry as well as of a potential in the matter couplings. In all cases investigated so
far, a strict separation of the potentials appearing in the geometric
part ($V$, $Z$ and $u$) and the one of the matter extension, $K$, occurred. 

From the result \eqref{eq:quant15}, non-local vertices of the matter fields can be
derived, which will be presented elsewhere. Another point to consider
is the definition of the remaining path integral measures and matter-loop
calculations. However, the model presented here is probably too
complicated for explicit calculations as the question turns out to be
very cumbersome in bosonic gravity with non-minimal coupling
\cite{Kummer:1997jr} and for supergravity with \emph{minimal} coupling
\cite{Bergamin:2004us} already.

Another line of investigations is the generalization to
extended supergravity \cite{Bergamin:2004sr} and/or to matter Lagrangians including
self-interaction. Here a better understanding of the result of this
work will be important. Indeed, the subtle interplay of the
constraints \eqref{eq:G++m}-\eqref{eq:G-m}, the structure functions
\eqref{eq:new3.1}-\eqref{eq:new4} and the Dirac brackets
\eqref{eq:dirbr1}-\eqref{eq:dirbr7} which led to the surprising result
of this section was not foreseeable. However, to keep the computational
complications manageable, exactly these important relations must be
understood a priori. Indeed, the results obtained so far suggest that
the first order formulation and the subsequent Hamiltonian
quantization provide a powerful symmetry principle, similar to the one
of superspace in the second order formulation. Nevertheless, in the
case of matter couplings the basics of this principle are not yet
understood sufficiently.

{\bf Acknowledgement.}  It is a pleasure to acknowledge interesting
discussions with D.~Grumiller,
W.~Kummer, D.~V.~Vassilevich and P.~van~Nieuwenhuizen. This work has been supported by the project
P-16030-N08 of the Austrian Science Foundation (FWF). I am especially
grateful to the organizers of GAS@BS and all participants for
the wonderful hospitality and numerous interesting conversations.

\appendix
\section{Notations and conventions}
\label{sec:notation}
These conventions are identical to
\cite{Ertl:2000si,Ertl:2001sj}, where additional explanations can be found.

The summation convention is $NW \rightarrow SE$ and almost every bosonic expression is transformed
trivially to the graded case when using this summation convention and
replacing commuting indices by general ones. This is possible together with
exterior derivatives acting \emph{from the right}, only. Thus the graded
Leibniz rule is given by $ \mbox{d}\left( AB\right) =A\mbox{d}B+\left( -1\right) ^{B}(\mbox{d}A) B$.

In terms of anholonomic indices the metric is $\mbox{diag}(1,-1)$ and the symplectic $2 \times 2$
tensors $\epsilon_{ab} = - \epsilon^{ab}$, $
\epsilon_{\alpha \beta} = \epsilon^{\alpha \beta}$ are defined with
$\epsilon_{01} = 1$. The metric in terms of holonomic indices is obtained by $g_{mn} = e_n^b e_m^a
\eta_{ab}$ and for the determinant the standard expression $e = \det e_m^a =
\sqrt{- \det g_{mn}}$ is used. The volume form reads $\epsilon = \half{1}
\epsilon^{ab} e_b \wedge e_a$; by definition $\ast \epsilon = 1$.

The $\gamma$-matrices are used in a chiral representation:
\begin{align}
\label{eq:gammadef}
  {{\gamma^0}_\alpha}^\beta &= \left( \begin{array}{cc} 0 & 1 \\ 1 & 0
  \end{array} \right) & {{\gamma^1}_\alpha}^\beta &= \left( \begin{array}{cc} 0 & 1 \\ -1 & 0
  \end{array} \right) & {{\gthree}_\alpha}^\beta &= {(\gamma^1
    \gamma^0)_\alpha}^\beta = \left( \begin{array}{cc} 1 & 0 \\ 0 & -1
  \end{array} \right)
\end{align}

Light-cone components are very convenient. As we work with spinors in a
chiral representation we can use $ \chi^\alpha = ( \chi^+, \chi^-)$
and upper and lower chiral components are related by $\chi^+
= \chi_-$, $ \chi^- = - \chi_+$, $\chi^2 = \chi^\alpha \chi_\alpha = 2 \chi_- \chi_+$. Vectors in light-cone coordinates are given by
$v^{++} = \frac{i}{\sqrt{2}} (v^0 + v^1)$, $v^{--} = \frac{-i}{\sqrt{2}}
  (v^0 - v^1)$.
The additional factor $i$ in permits a direct identification of the light-cone components with
the components of the spin-tensor $v^{\alpha \beta} = \frac{i}{\sqrt{2}} v^c \gamma_c^{\alpha
  \beta}$. This implies that $\eta_{++|--} = 1$
and $\epsilon_{--|++} = - \epsilon_{++|--} = 1$. In light-cone coordinates each
$\gamma$-matrix has exactly one non-vanishing entry, namely $
{(\gamma^{++})_+}^- = \sqrt{2} i$, $
{(\gamma^{--})_-}^+ = -\sqrt{2} i$.


\begin{thebibliography}{9}
\small

\bibitem{Bergamin:2004us}
L.~Bergamin, D.~Grumiller, and W.~Kummer.
\newblock Quantization of 2d dilaton supergravity with matter.
\newblock {\em JHEP}, 05:060, 2004.

\bibitem{Bergamin:2003mh}
L.~Bergamin, D.~Grumiller, and W.~Kummer.
\newblock Supersymmetric black holes in 2-d dilaton supergravity: baldness and
  extremality.
\newblock {\em J. Phys.}, A37:3881--3901, 2004.

\bibitem{Bergamin:2003am}
L.~Bergamin and W.~Kummer.
\newblock The complete solution of 2d superfield supergravity from graded
  poisson-sigma models and the super pointparticle.
\newblock {\em Phys. Rev.}, D68:104005, 2003.

\bibitem{Bergamin:2002ju}
L.~Bergamin and W.~Kummer.
\newblock Graded poisson sigma models and dilaton-deformed 2d supergravity
  algebra.
\newblock {\em JHEP}, 05:074, 2003.

\bibitem{Bergamin:2004sr}
L.~Bergamin and W.~Kummer.
\newblock Two-dimensional {N}=(2,2) dilaton supergravity from graded
  poisson-sigma models.
\newblock 2004.

\bibitem{Dirac:1996}
P.~A.~M. Dirac.
\newblock {\em {Lectures on Quantum Mechanics}}.
\newblock Belfer Graduate School of Science, Yeshiva University, New York,
  1996.

\bibitem{Ertl:2000si}
M.~Ertl, W.~Kummer, and T.~Strobl.
\newblock General two-dimensional supergravity from {P}oisson superalgebras.
\newblock {\em JHEP}, 01:042, 2001.

\bibitem{Ertl:2001sj}
Martin Ertl.
\newblock {\em Supergravity in two spacetime dimensions}.
\newblock PhD thesis, {T}echnische {U}niversit{\"a}t {W}ien, 2001.

\bibitem{Grumiller:2003sk}
D.~Grumiller and W.~Kummer.
\newblock How to approach quantum gravity: Background independence in 1+1
  dimensions.
\newblock In N.~M. Borstnik, H.~B. Nielsen, C.~D. Froggat, and D.~Lukman,
  editors, {\em Proceedings to the {E}uroconference on symmetries beyond the
  {S}tandard {Model}}, volume~4 of {\em Bled Workshops in Physics}, pages
  184--196, 2003.

\bibitem{Grumiller:2002nm}
D.~Grumiller, W.~Kummer, and D.~V. Vassilevich.
\newblock Dilaton gravity in two dimensions.
\newblock {\em Phys. Rept.}, 369:327, 2002.

\bibitem{Howe:1979ia}
P.~S. Howe.
\newblock Super {W}eyl transformations in two-dimensions.
\newblock {\em J. Phys.}, A12:393--402, 1979.

\bibitem{Izquierdo:1998hg}
J.~M. Izquierdo.
\newblock Free differential algebras and generic 2d dilatonic (super)gravities.
\newblock {\em Phys. Rev.}, D59:084017, 1999.

\bibitem{Katanaev:1996bh}
M.~O. Katanaev, W.~Kummer, and H.~Liebl.
\newblock {Geometric Interpretation and Classification of Global Solutions in
  Generalized Dilaton Gravity}.
\newblock {\em Phys. Rev.}, D53:5609--5618, 1996.

\bibitem{Kummer:1997jr}
W.~Kummer, H.~Liebl, and D.~V. Vassilevich.
\newblock Hawking radiation for non-minimally coupled matter from generalized
  2d black hole models.
\newblock {\em Mod. Phys. Lett.}, A12:2683--2690, 1997.

\bibitem{Park:1993sd}
Young-Chul Park and Andrew Strominger.
\newblock Supersymmetry and positive energy in classical and quantum
  two-dimensional dilaton gravity.
\newblock {\em Phys. Rev.}, D47:1569--1575, 1993.

\bibitem{Schaller:1994uj}
Peter Schaller and Thomas Strobl.
\newblock {Poisson sigma models: A generalization of 2-d gravity Yang- Mills
  systems}.
\newblock In {\em Finite dimensional integrable systems}, pages 181--190, 1994.
\newblock Dubna.

\bibitem{Schaller:1994es}
Peter Schaller and Thomas Strobl.
\newblock Poisson structure induced (topological) field theories.
\newblock {\em Mod. Phys. Lett.}, A9:3129--3136, 1994.

\end{thebibliography}
\end{document}